\documentclass[runningheads]{llncs}

\usepackage[T1]{fontenc}
\usepackage{graphicx}
\usepackage{booktabs}
\usepackage{amsmath}
\usepackage{xcolor}
\usepackage{multirow}
\usepackage{diagbox}
\usepackage{dirtytalk}
\usepackage{colortbl}
\usepackage[most]{tcolorbox}
\usepackage{enumitem}
\usepackage{xurl} 
\usepackage{cite} 
\usepackage{csquotes}
\usepackage{listings}
\usepackage[title]{appendix}
\usepackage{tablefootnote}
\usepackage{algorithm}
\usepackage{algpseudocode}
\usepackage{epigraph}
\usepackage{booktabs}
\usepackage{float}
\usepackage[depth=3]{bookmark}
\usepackage{minitoc}
\usepackage{orcidlink}
\usepackage{pifont}

\definecolor{color_Imitation}{HTML}{893244}
\definecolor{color_Translation}{HTML}{4F404C}
\definecolor{color_Obfuscation}{HTML}{007F7F}
\definecolor{color_Steganography}{HTML}{6082B6}
\definecolor{color_Soundness}{HTML}{00674F}
\definecolor{color_Safety}{HTML}{ffbf00}
\definecolor{color_Sensibility}{HTML}{819171}
\definecolor{color_Adversarial_Stylometry}{HTML}{971a1e}
\definecolor{color_a}{HTML}{893244}
\definecolor{color_b}{HTML}{4F404C}
\definecolor{color_c}{HTML}{007F7F}
\definecolor{color_d}{HTML}{6082B6}
\newcommand{\myImitation}{\textcolor{color_Imitation}{Imitation}}
\newcommand{\myTranslation}{\textcolor{color_Translation}{Translation}}
\newcommand{\myObfuscation}{\textcolor{color_Obfuscation}{Obfuscation}}
\newcommand{\mySteganography}{\textcolor{color_Steganography}{Steganography}}
\newcommand{\mySoundness}{\textcolor{color_Soundness}{Soundness}}
\newcommand{\mySafety}{\textcolor{color_Safety}{Safety}}
\newcommand{\mySensibility}{\textcolor{color_Sensibility}{Sensibility}}
\newcommand{\myAdversarial}{\textcolor{color_Adversarial_Stylometry}{Adversarial}}
\newcommand{\myPlus}{\( \ + \ \)}

\definecolor{color_Additional_01}{HTML}{f9f9f9}
\definecolor{color_Additional_02}{HTML}{a6a6ed}

\lstdefinestyle{mystyle}{   
    backgroundcolor=\color{color_Additional_01},
    commentstyle=\color{color_Obfuscation},
    keywordstyle=\color{color_Additional_02},
    numberstyle=\tiny\color{color_Sensibility},
    stringstyle=\color{color_Safety},
    basicstyle=\ttfamily\footnotesize,
    breakatwhitespace=false,         
    breaklines=true,                 
    captionpos=b,                    
    keepspaces=true,                 
    numbers=left,                    
    numbersep=5pt,                  
    showspaces=false,                
    showstringspaces=false,
    showtabs=false,                  
    tabsize=2
}

\title{
    Unveiling Unicode's Unseen Underpinnings in Undermining Authorship Attribution
}

\author{
    Robert Dilworth \orcidlink{0009-0005-5497-9810}
}

\authorrunning{
    Robert Dilworth
}
\titlerunning{
    Adversarial Stylometry Embedded Steganographically
}

\institute{
    Department of Computer Science and Engineering, Mississippi State University, Mississippi State, Mississippi, USA\\
    \email{rkd103@msstate.edu}
}

\hypersetup{
    pdftitle={Unveiling Unicode's Unseen Underpinnings in Undermining Authorship Attribution},
    pdfsubject={cs.CR, cs.CL, cs.IR},
    pdfauthor={Robert Dilworth},
    pdfkeywords={Unicode Steganography with Zero-Width Characters, Adversarial Stylometry, Privacy},
    colorlinks=true,
    linkcolor=color_a,
    citecolor=color_c,    
    urlcolor=color_d,
}

\begin{document}

\maketitle

\begin{abstract}

    When using a public communication channel---whether formal or informal, such as commenting or posting on social media---end users have no expectation of privacy: they compose a message and broadcast it for the world to see. Even if an end user takes utmost precautions to anonymize their online presence---using an alias or pseudonym; masking their IP address; spoofing their geolocation; concealing their operating system and user agent; deploying encryption; registering with a disposable phone number or email; disabling non-essential settings; revoking permissions; and blocking cookies and fingerprinting---one obvious element still lingers: the message itself. Assuming they avoid lapses in judgment or accidental self-exposure, there should be little evidence to validate their actual identity, right? Wrong. The content of their message---necessarily open for public consumption---exposes an attack vector: stylometric analysis, or author profiling. In this paper, we dissect the technique of stylometry, discuss an antithetical counter-strategy in adversarial stylometry, and devise enhancements through Unicode steganography.

    
\end{abstract}

\section{Introduction}
\label{sec:Introduction}

    \textit{Steganography} and \textit{stylometry} \cite{Gupta2019, Savoy2020} are two sides of the same coin. While steganography---the concealing of data in innocuous files---is employed to \textit{elude} detection, stylometry serves to \textit{bolster} detection. Granted, the domains (or problem spaces) in which they are applied typically differ. For instance, ponder the following. Do you want to convey a message while skirting the watchful eyes of onlookers? Then, steganography is your best bet. Do you need to profile the writer of a message, unearthing the author's demographics---gender, age, native language(s) (vernacular features), level of education, and ethnicity? In that case, stylometry is the solution. In this way, if stylometry \textit{compromises} privacy, then steganography could conceivably \textit{enhance} it, but that view is \textit{myopic}. 

    Drawing from this narrow-minded perspective, a saying comes to mind: \say{the best offense is a good defense.} Namely, by developing a deep and nuanced understanding of your adversary, you can better craft a more tailored, bespoke defense. Thus, if the goal is to preserve privacy, it naturally follows that one would need to comprehend the \textit{tools} that can impede the objective of remaining undetectable. That's where \textit{adversarial stylometry} \cite{Juola2012,McDonald2012,Almishari2014,Zhai2022,Saedi2020,Bevendorff2019,Haroon2021,Gröndahl2020,Adelani2021,Wang2022,Neal2017,Uchendu2023,Mireshghallah2021,Xu2019,Potthast2016,Afroz2012,Mahmood2019,Shetty2018,Emmery2021,Narayanan2012,Gröndahl2019,Juola2011,Brennan2009,Kacmarcik2006,Brennan2012,Rao2000,Xing2024} comes into play---flipping the script by recasting \say{the enemy of your enemy is your ally} into a counterintuitive strategy of misdirection.
    
    If embedding media within media falls short, then let's layer the \textit{imperceptible} with the \textit{perceptible} to throw off the scent of discovery. What if we apply \textit{antagonistic} stylometric principles to the media that will eventually house embedded content? Would that favorably or adversely impact stylometric detection? Intuitively, the saying \say{the more the merrier} seems applicable here; however, our hypothesis's veracity remains untested. We now proceed to articulate our problem statement more clearly.

    \subsection{Problem Statement}
    \label{subsec:Problem_Statement}

        \begin{tcolorbox}[colback = gray!20!white, colframe = gray!75!black]
            
            If we take some digitized media---for the sake of argument, raw text or textual representations of non-textual forms of media---and inject another piece of media into the original while obscuring the embedding by either (A) imitating the idiosyncratic writing style of another author, (B) executing multiple rounds of machine translation from various dissimilar languages, or (C) obscuring the original style, whether manually or automatically, or some \textit{permutative} combination of (A), (B), and (C), will the \textit{grammatical} integrity, \textit{syntactical} structure, and \textit{semantical} meaning of the original remain sufficiently intact to avoid detection by a stylometric system?
            
        \end{tcolorbox}

\begin{figure}[H]
            \centering
            \includegraphics[width=1\linewidth]{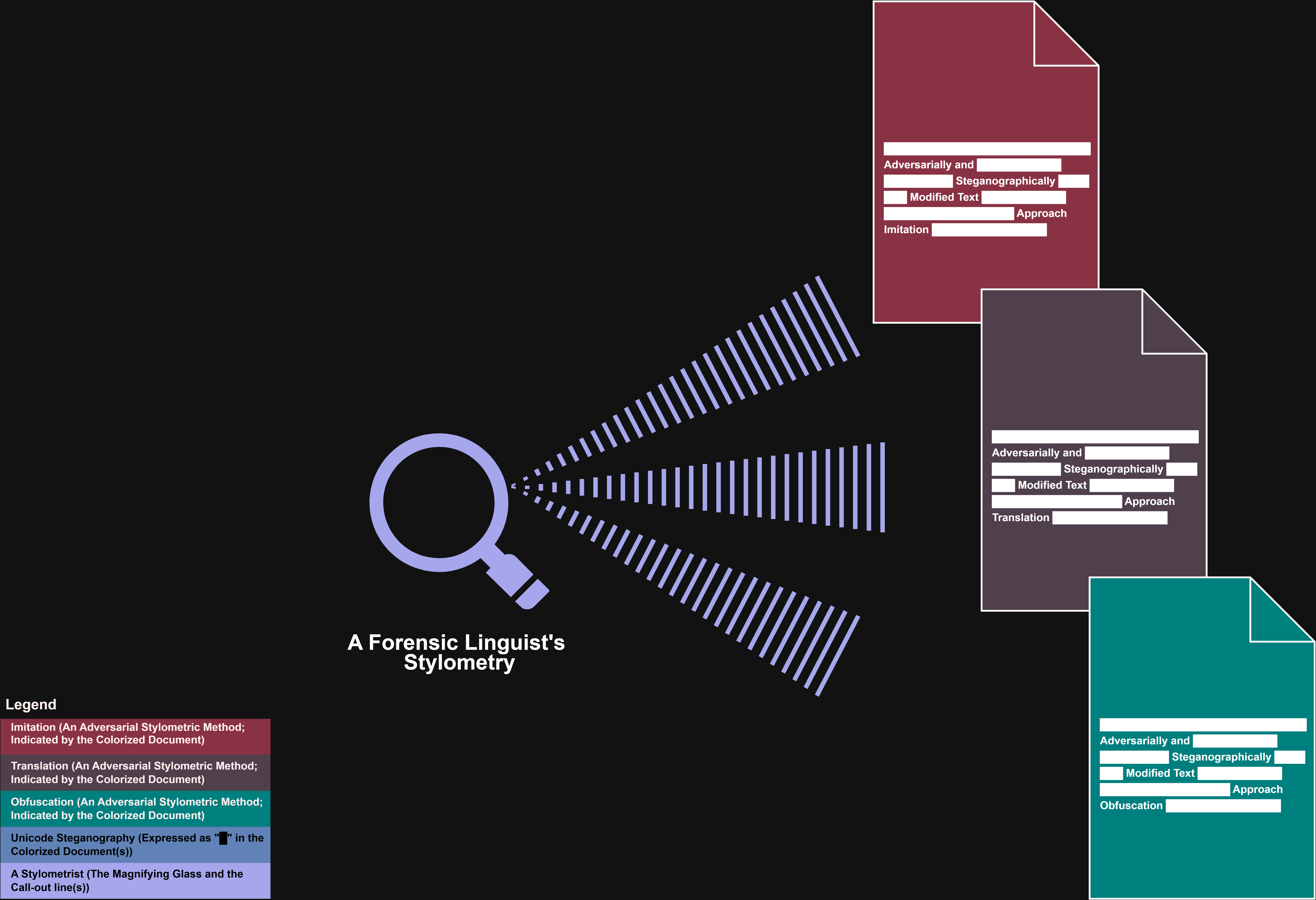}
            \caption{Overview of Adversarial Stylometric Methods (\myImitation, \myTranslation, \myObfuscation, Unicode \mySteganography?) and Stylometrist Examination}
            \label{fig:Adversarial_Stylometry_Overview}
        \end{figure}

        Building on this question of media manipulation and covert transformation, we seek to circumvent stylometric detection via technical \textit{subterfuge} and \textit{skulduggery} by injecting steganographic content, whether sensible or nonsensical, into adversarially modified text. For our intents and purposes, the incorporation of steganography is \say{a means to an end, not an end in itself;} future experimentation will ascertain the effect of steganographic embeddings \textit{vis-\`a-vis} various forms of stylometric analysis. See (\textbf{Figure \ref{fig:Adversarial_Stylometry_Overview}}) for an overview.

    \subsection{Paper Structure}
    \label{subsec:Paper_Structure}
    
        That, (\textbf{Section \ref{subsec:Problem_Statement}}), is the question that will guide the formulation of this paper. First, we will motivate the study by debating the use of stylometry, whether from an adversarial or an ally perspective (\textbf{Section \ref{sec:Dilemma_Adversarial_Stylometry}}). Then, we clarify the specific flavor of steganography that we wish to test: the variant that utilizes Unicode characters that, when properly rendered, are imperceptible to the naked eye (\textbf{Section \ref{sec:Conceptual_Overview}}). Next, we logically unpack our hypothesis by weighing the pros and cons of a dual steganography/adversarial stylometry framework (\textbf{Sections \ref{sec:Zero-Width_Adversarial_Stylometry}; \ref{sec:Potential_Benefits_Limitations_and_Challenges}}). Then, we discuss the various modes of adversarial stylometry and the metrics used to evaluate them (\textbf{Sections \ref{sec:Combining_Steganography_Imitation_Translation_Obfuscation}; \ref{sec:Evaluation_Metrics_Combined_Approach}; \ref{sec:Strategic_Considerations}}). Throughout, we will interject our musings with the ways in which our hypothesized system could be exploited: consider a world in which one can almost perfectly evade stylometric detection and the consequences of never being able to definitively identify an author or verify authorship (\textbf{Section \ref{sec:Real-World_Considerations}}). Penultimately, we plan our prospective investigative pursuit, including but not limited to sketching a preliminary strategy and procuring a dataset (\textbf{Section \ref{sec:Future_Work}}). Finally, we conclude by highlighting the potential positive impact on privacy, where the goal is to \textit{disclose the least amount of information---overtly or covertly---to the fewest people (or machines or agentic observers) possible} (\textbf{Section \ref{sec:Conclusion}}).

\section{The Dilemma of Adversarial Stylometry: Exploring Arguments and Implications}
\label{sec:Dilemma_Adversarial_Stylometry}

    \subsection{A Case for Stylometry}
    \label{subsec:Positive_Case_Stylometry}

        Place yourself in the following scenario. You have a sibling who never fit in growing up, but they \textit{were} exceptionally bright---so much so that they would eventually go on to obtain a terminal degree. After they acquired a well-regarded profession, their psyche and cognitive reserves, regrettably, began to silently and slowly erode. 
        
        They entertained \textit{radical} thoughts and subsisted on media that further solidified their \textit{warped} worldview. One day, one seemingly inconsequential event transpired, serving as the catalyst that gradually \textit{goaded} their descent into \textit{depravity}. They became \textit{irate} at the world and society at large for not only failing them but also trending toward its collective ruin. 
        
        Their \textit{ire} evolved into \textit{misanthropy}, and their \textit{misanthropy} instilled within them an unmatched \textit{schadenfreude}. You, as their loving sibling, tried to support them as best as you could, but they continued to conceal their inner turmoil and retreat into reclusion. You figured everything would work out and things would resolve themselves with time; your sibling was brilliant after all. 
        
        However, after years of sparse, intermittent contact, you stumble upon a manifesto online detailing how humanity was intrinsically \textit{morally reprehensible} and collectively deserving of \textit{perdition} and \textit{annihilation}. You initially scoff at the work, chalking it up as the \textit{abstruse} ramblings of a disturbed individual, but your inquisitiveness gets the better of you and you read the manuscript cover to cover. 
        
        As you turn each page, a deep, sinking feeling settles in your stomach. \say{No, it can't be;} but, indeed, you are almost certain. The way that the author describes their thoughts, the logical flow of their ideas, the prose rife with \textit{stream of consciousness}, the peculiar choice of words, and the unnerving and repeated use of \textit{assonance} all but confirm your fears. Your sibling and the \textit{eccentric} author are one and the same. Worst yet, you discover that your sibling is an infamous \say{ideologically motivated insurgent} who has maimed and brutalized innocents. 
        
        You, appalled by their behavior and remorseful for what you must do, reluctantly notify the authorities. Your decision to sell out the culprit eventually saved lives, \textit{yes}, but it came at the cost of losing your one and only sibling. 
        
        Now, consider this: if your sibling wrote a 300-page manifesto, it would take a non-trivial amount of time to parse through their wrath-laden musings. Nevertheless, that's where stylometry could enter the picture. Assuming the manifesto was typed (or digitally transcribed---using OCR\footnote{Optical character recognition.} to convert scanned pages into editable text), it could be efficiently parsed and fed into a machine learning algorithm to glean various insights, such as the author's writing style. 
        
        Armed with these insights, you, the sibling of the \say{homicide perpetrator,} could then compare your sibling's manifesto against their other scholarly or non-scholarly publications, like their dissertation or personal blog. In an alternate timeline where you made use of stylometry---and a strong correlation between your sibling and the \say{ideologically motivated insurgent's} writings was revealed---further casualties could have been prevented by eliminating the manual effort of reading and taking notes, replacing the laborious task with an automated, machine learning workflow. 
        
        In this instance, removing your sibling's \textit{anonymity} (note: we neglected to mention that your sibling penned their manifesto under a \textit{pseudonym}) and invading their \textit{privacy} would lead to a net societal good.\footnote{As a disclaimer, creative liberties were taken as we fictionalized the true account of the Unabomber incident as recounted by David Kaczynski in \textit{Every Last Tie: The Story of the Unabomber and His Family} \cite{Kaczynski2015}.}
        
    \subsection{A Case Against Stylometry: Necessitating the Need for Adversarial Stylometry}
    \label{subsec:Negative_Case_Stylometry}
        
        Picture this. A \textit{beleaguered} citizen of an \textit{authoritarian} regime, weighed down by the daily oppressive realities, musters the courage to anonymously criticize and actively revolt against their \textit{corrupt}, \textit{morally debunk} dictator.
        
        The regime, seeking to \say{nip the seeds of rebellion in the bud,} employs stylometry to make a reasonable determination of the \textit{slanderous} material's author. The author---in the absence of \textit{anonymity} and \textit{basic inalienable rights}---mysteriously vanishes\footnote{\say{Disappeared} in the \say{abducted} and \say{whereabouts concealed} sort of way.} without a trace as the regime further \textit{putrefies} and \textit{decays}. 
        
        In this instance, it would be wise to take every possible measure to anonymize a message, be it \textit{incendiary} or otherwise, by using a combination of privacy-oriented mechanisms like \textit{onion routing} (Tor\footnote{\url{https://www.torproject.org/}}) and \textit{virtual private networks} (VPNs\footnote{\url{https://protonvpn.com/}}). Applying an added layer of stylometric-thwarting methods (like adversarial stylometry) to a \textit{VPN-obfuscated}, \textit{multi-layer encrypted}, \textit{poly-node-relayed} connection would be the safest course of action in such a scenario. Indeed, if you're at risk of \say{erasure} for freely speaking your mind (like an indefinite prison sentence or, worst yet, your untimely demise), such cloaking measures are required, \textit{nay}, necessary.
        
    \subsection{Closing Remarks: To ``Stylometry," or Not to ``Stylometry," That is the Question}
    \label{subsec:Closing_Remarks_Stylometry}
        
        \textit{Cybersecurity} is a never-ending game of \say{cat and mouse.} Much like the vintage cartoon \textit{Tom and Jerry.} One moment Tom (think of the feline as a cyber defender) has the upper hand, only to be outwitted and outmaneuvered by the pesky rodent, Jerry (think of the mouse as a malicious actor). Jerry, based on how they're presented in the show, is elusive and resourceful; unless they want to be seen, they will remain enigmatic and anonymous. Tom, too, needs to remain covert in their operations so that they can gain the upper hand against their opponent. While both parties are \textit{diametrically} opposed, they both recognize the utility of camouflage (think of their comical, convoluted \textit{hijinks} as adversarial stylometry). 
        
        As we step outside the nostalgic analogy, the need for research into adversarial stylometry should be evident from the previously provided accounts. Stylometry can be deployed not only to \textit{save lives} but also to \textit{threaten} them. Furthermore---as you will hopefully come to appreciate---steganography could also be deployed to diminish stylometric measures. This interplay between life-saving and life-threatening applications sets the stage for a deeper exploration of the underlying concepts, which we detail in the overview below.

\section{Overview of Key Concepts}
\label{sec:Conceptual_Overview}

    \begin{enumerate}
    
        \item[\ding{118}] \textbf{Unicode Steganography with Zero-Width Characters (\textit{Zaynalov et al.} \cite{Zaynalov2020} \& \textit{Thompson} \cite{Thompson2021}):}
        
            \begin{itemize}
            
                \item Zero-width characters (like \textit{Zero-Width Space} \texttt{[U+200B]}, \textit{Zero-Width Non-Joiner} \texttt{[U+200C]}, \textit{Zero-Width Joiner} \texttt{[U+200D]}, and \textit{Zero-Width No-Break Space} \texttt{[U+FEFF]}) are invisible in rendered text.
                \item They can be inserted into text without altering its visible appearance, effectively hiding additional information within a message.
                \item This hidden data can encode \textit{metadata} or signals that might otherwise be recoverable only by sophisticated processing.
                \item In a hypothetical encoding scheme, a Zero-Width Space could represent a \say{0} (the absence of a signal) while a Zero-Width Non-Joiner could represent a \say{1} (the presence of a signal), enabling the binary representation of information. Any of the remaining zero-width characters---Zero-Width Joiner or Zero-Width No-Break Space---could serve as the other essential tokens: one to delimit \textit{lexemes} (letters or words in this case) and one to mark the end of a line. See (\textbf{Appendices \ref{appx:Steganography_Python_Example}; \ref{appx:Steganography_Pseudocode}}).
                \item One method of detecting Unicode steganography is to view a text file in a hexadecimal (hex) editor, which reveals the raw byte stream and allows you to spot unexpected code points and steganographic payloads.
                
            \end{itemize}
    
        \item[\ding{118}] \textbf{Adversarial Stylometry (\textit{Rao et al.} \cite{Rao2000}):}
        
            \begin{itemize}
            
                \item Stylometry analyzes writing style to attribute authorship by examining features such as vocabulary, syntax, punctuation, and even invisible formatting nuances.
                \item A closely intertwined, basic application is the verification problem, which determines whether a supplied text is produced by a given set of authors---be it a single candidate or a reasonably sized list of candidates. Nesting an array of verification problems with binary, yes-or-no responses constitutes the \textit{authorial attribution} for which stylometry is best known.
                \item Adversarial stylometry involves deliberately altering or obfuscating these stylistic features to evade or mislead authorship attribution algorithms.
                \item Attacks in this space might include subtle modifications that do not change the semantic meaning of the text but interfere with feature extraction.
                
            \end{itemize}
        
    \end{enumerate}

    With the key terms defined, we now examine how zero-width characters can supplement adversarial stylometry.

\section{How Zero-Width Characters Can Aid Adversarial Stylometry}
\label{sec:Zero-Width_Adversarial_Stylometry}

    \subsection{Embedding Noise or Decoys}
    \label{subsec:Embedding_Noise_or_Decoys}

        By inserting zero-width characters at calculated places (for instance, between words, at sentence boundaries, or even within words), an adversary can introduce noise that may obscure the traditional stylistic markers. This type of \say{hidden noise} can affect statistical profiles that stylometric algorithms build, potentially leading to \textit{misattribution} or even reducing confidence scores.

    \subsection{Signal Encoding and Feature Diversion}
    \label{subsec:Signal_Encoding_and_Feature_Diversion}
    
        One can hide carefully crafted signals within the text that may change the \textit{parser's} output when the text is analyzed. For example, an adversary might encode bits that correspond to extra token boundaries or influence \textit{tokenization} in a way that artificial features appear, thus \textit{distorting} the author's genuine stylistic profile.

    \subsection{Creating Multiple ``Layers'' of Style}
    \label{subsec:Creating_Multiple_Layers_of_Style}

        The visible, tampered text retains the intended human-readable style while the embedded zero-width characters add another layer that conventional stylometric tools might \textit{inadvertently} process if not filtered out. Such a dual-layer approach can lead to adversaries controlling which stylistic signals are \say{seen} by automated methods without affecting human \textit{perception}.
        
        By mastering the placement and interpretation of these hidden markers, an adversary not only \textit{hijacks} another's stylistic fingerprint but also reasserts \textit{ultimate authority} over the text's use and identity---thus leading us to the core principle:

            \begin{displayquote}
    
                When all is said and done, you only truly \textit{own} something to the extent that you can \textit{control} said thing, and the willful inclusion or exclusion of material---irrespective of the owner's incentive---is a \textit{proprietor's} right, doubly so when dealing with data. And if safeguarding that data means \say{poisoning your own well,} so be it: your data, your poison.
    
            \end{displayquote}

    Having explored how zero-width characters can enhance adversarial stylometry, we now turn our attention to the tradeoffs of this approach.

\section{Potential Benefits, Limitations, and Challenges}
\label{sec:Potential_Benefits_Limitations_and_Challenges}

    \begin{itemize}

        \item \textbf{Merits}:

            \begin{itemize}
            
                \item \textit{Increased Robustness Against Attribution}: Adding invisible mutations can hinder the stable extraction of features needed for author identification.
                    
                \item \textit{Flexibility}: Zero-width characters allow for subtle, nearly undetectable modifications that can be tailored based on the target attribution algorithm.
                    
                \item \textit{Reversibility and Selectivity}: In some steganographic schemes, the alterations might be reversible for authorized users but still confusing to third-party analysis systems. 
                
            \end{itemize}

        \item \textbf{Drawbacks:}

            \begin{itemize}
            
                \item \textit{Detection Mechanisms}: Advanced stylometric tools might incorporate preprocessing steps to strip out zero-width characters. If these characters are noted as unusual, detection methods may improve.
                    
                \item \textit{Transfer and Rendering}: Not all text-rendering systems or processing pipelines preserve zero-width characters. Their removal or transformation may \textit{negate} the hidden modifications.
                    
                \item \textit{Unintended Statistical Artifacts}: While the goal is to obfuscate, inadvertently creating statistical \textit{outliers} might further flag texts as manipulated or serve as an identifying \say{fingerprint} of adversarial intervention.
                    
                \item \textit{Consistency}: The technique requires careful calibration to ensure that the modifications do not interfere with the overall readability or natural flow when processed by natural language algorithms.
                
            \end{itemize}
        
    \end{itemize}

    While the previous section explored the \textit{theoretical} landscape---highlighting various considerations---it now becomes imperative to ground these ideas in \textit{practical} reality. In moving forward, we delve into how these concepts translate into actionable strategies, the obstacles they may encounter in everyday applications, and the implications for actual implementation.

\section{Real-World Considerations}
\label{sec:Real-World_Considerations}

    Research in \textit{adversarial machine learning} is ongoing, and methods like these are likely to trigger a counter-reaction from \textit{forensic linguistics} researchers who develop more robust de-stylometry methods. The adversarial benefits need to be weighed against the possibility of inadvertently embedding detectable patterns that could become forensic artifacts themselves. 
    
    Ethical and legal ramifications must also be considered when deploying such techniques, especially in contexts like \textit{academic integrity} or if used to obfuscate \textit{unauthorized authorship} or \textit{malicious content}.

    We now broaden our discourse by addressing the tangible facets of adversarial stylometry usage, hashing out \say{the good, the bad, and the ugly.}

    \subsection{Discussion}
    \label{subsec:Discussion}

        In a world where stylometric detection can be almost perfectly evaded, the very boundaries of authorship and identity become fluid, dissolving into a \textit{murk} of mystery and ambiguity. In such a domain, the written word---a tool once considered \textit{indelibly} marked by its creator---loses its power to be tied to a singular, accountable hand. The specter of every writer haunts the text, a collective, nebulous \textit{umbra} that defies clear attribution.

        This scenario forces us to confront an existential paradox: while anonymity might \textit{embolden} free expression and protect vulnerable voices from persecution, it simultaneously \textit{robs} literature of its lineage. When every sentence could be the creation of many, or none at all, the trust we place in words begins to waver. Authorship, traditionally a badge of honor and responsibility, transforms into a relic of the past---an artifact whose authenticity is perpetually up for debate.
        
        Philosophically, the implications are significant. The removal of definitive authorship challenges our understanding of \textit{creativity} and \textit{originality}. If texts can exist without a detectable creator, do we begin to see them as autonomous entities, evolving and interacting beyond the confines of their initial conception? This prompts a re-evaluation of \textit{intellectual property} and the very nature of \textit{artistic expression}: who owns a work when its provenance is indeterminate?
        
        Furthermore, the erosion of fixed identity in written expression raises questions about \textit{accountability}. Literature and communication are not merely about \textit{aesthetics} or \textit{function}; they are also instruments of \textit{responsibility}. In a society where texts---and by extension, their creators---cannot be held accountable, there lies the potential for both unprecedented \textit{liberation} and profound \textit{pandemonium}. \textit{Truth} becomes \textit{elusive}, and the foundations of \textit{trust} in discourse \textit{retrograde}---sliding backward into \textit{obsolescence}---further entangling the web of human interactions.
        
        Yet, there exists a compelling counterargument. In an era where \textit{telemetry abuse} and the forceful collection of data points can coalesce into a comprehensive \textit{dossier} on an individual, preserving privacy at all costs is synonymous with reclaiming personal \textit{agency}. By intentionally \textit{fragmenting} and \textit{obfuscating} one's digital \textit{persona}, one can resist the \textit{invasive} tendencies of algorithmic \textit{profiling} and ad ecosystems. Rather than tailoring our behavior to fit predictable models, we could purposely create a \textit{shifting}, \textit{untrackable}, \textit{unpredictable} presence---rejecting the notion of a \textit{consistent} and \textit{persistent} digital identity. This disciplined approach to online conduct, from \textit{blocking} scripts to \textit{poisoning} data profiles to \textit{tarnishing} knowledge graphs to \textit{hardening} machinery, isn't about \textit{erasing} one's existence but about \textit{safeguarding} one's autonomy. In a landscape where every click feeds into the \textit{cogs} of surveillance capitalism, maintaining a \textit{phantasmically ephemeral}, \textit{noise-filled} presence is the ultimate act of \textit{self-defense}.

        \begin{tcolorbox}[colback = gray!20!white, colframe = gray!75!black]
            
            When \say{the walls have ears} (\textit{ever listening\dots}) and the \textit{panopticon's} disembodied eyes are innumerable and unyielding (\textit{ever watching\dots}), it's not about what one has to \textit{hide}, but rather what one must \textit{protect}.

        \end{tcolorbox}
        
        Ultimately, a world in which stylometric detection fails to pinpoint authorship compels us to rethink fundamental concepts of originality, accountability, and individual expression. It presents not only a technical challenge but also a profound philosophical probe into the nature of \textit{identity}, \textit{creativity}, and \textit{control} in the digital age.

\section{Combining Steganography with Imitation, Translation, and Obfuscation}
\label{sec:Combining_Steganography_Imitation_Translation_Obfuscation}

    Following our exploration of the philosophical implications of non-definitive authorship and fractured identity, we now transition to a pragmatic examination of adversarial stylometric strategy. A multi-layered adversarial approach can benefit from blending multiple strategies, such as imitation, translation, and obfuscation: classical techniques pioneered by \textit{Neal et al.} \cite{Neal2017}.
    
    \subsection{Imitation}
    \label{subsec:Imitation}
        
        \textit{Imitation} involves mimicking the stylistic features of another author or a generic style that is less distinctive. When paired with zero-width steganography, one could secondarily encode decoy signals that reflect the target style. This may involve intentionally choosing punctuation, syntax, or vocabulary that mimics a reference dataset, while the hidden characters further obscure the original style. See (\textbf{Figure \ref{fig:Six_Paths_of_Transformation}}) for an example.\footnote{\url{https://www.rejectconvenience.com/privacy-visualizer/}}
    
    \subsection{Translation}
    \label{subsec:Translation}
    
        Machine or human \textit{translation} offers an avenue for breaking some of the inherent stylistic fingerprints of an author's native tongue. After translating the content to another language and then back (or to multiple languages in a chain), the stylistic markers become less reliable. Embedding zero-width characters on top of the translated text can help control feature extraction, ensuring that these latent signals guide analysis away from the original style. See (\textbf{Figure \ref{fig:Six_Paths_of_Transformation}}) for an example.
    
    \subsection{Obfuscation}
    \label{subsec:Obfuscation}
        
        Traditional \textit{obfuscation} techniques involve deliberately altering or suppressing certain stylistic markers, like randomizing or shuffling elements within the text. Zero-width steganography, in this context, can serve as a covert channel to either inject or remove signals in \textit{synchrony} with visible obfuscation. This could produce a two-tiered obfuscation where both \textit{overt} (visible) and \textit{covert} (invisible) modifications attempt to \textit{befuddle} stylometric attribution. See (\textbf{Figure \ref{fig:Six_Paths_of_Transformation}}) for an example.

\section{Evaluation Metrics for the Combined Approach}
\label{sec:Evaluation_Metrics_Combined_Approach}

    A rigorous evaluation of the combined methods (imitation, translation, and obfuscation) should address the criteria of soundness, safety, and sensitivity as articulated by \textit{Potthast et al.} \cite{Potthast2016}.

    \subsection{Soundness}
    \label{subsec:Soundness}

        \textit{Soundness} refers to whether the modifications (both overt and covert) maintain the text's \textit{integrity}, \textit{meaning}, and \textit{readability}. The approach should ensure that while hidden characters and altered stylistic features \textit{mislead} stylometric systems, they do not \textit{disrupt} the semantic content or lead to outright syntactical errors. Evaluation should consider the \textit{fidelity} of the message and whether the modifications can be reversed or recognized (by authorized parties) without \textit{degradation}. See (\textbf{Figure \ref{fig:Six_Paths_of_Transformation}}) for an example.

    \subsection{Safety}
    \label{subsec:Safety}

        \textit{Safety} involves the risk of detection and potential collateral issues, such as the technique inadvertently creating artifacts that forensic tools might exploit to identify manipulated texts. The combined approach must minimize \textit{side effects}, like patterns easily detectable by enhanced preprocessing filters that remove zero-width characters. Safety also covers any unintended legal or ethical implications, particularly if the obfuscation is used in contexts like \textit{plagiarism}, \textit{misinformation}, or other \textit{malicious activities}. See (\textbf{Figure \ref{fig:Six_Paths_of_Transformation}}) for an example.

    \subsection{Sensibility}
    \label{subsec:Sensibility}

\begin{figure}[H]
        \centering
        \includegraphics[width=1\linewidth]{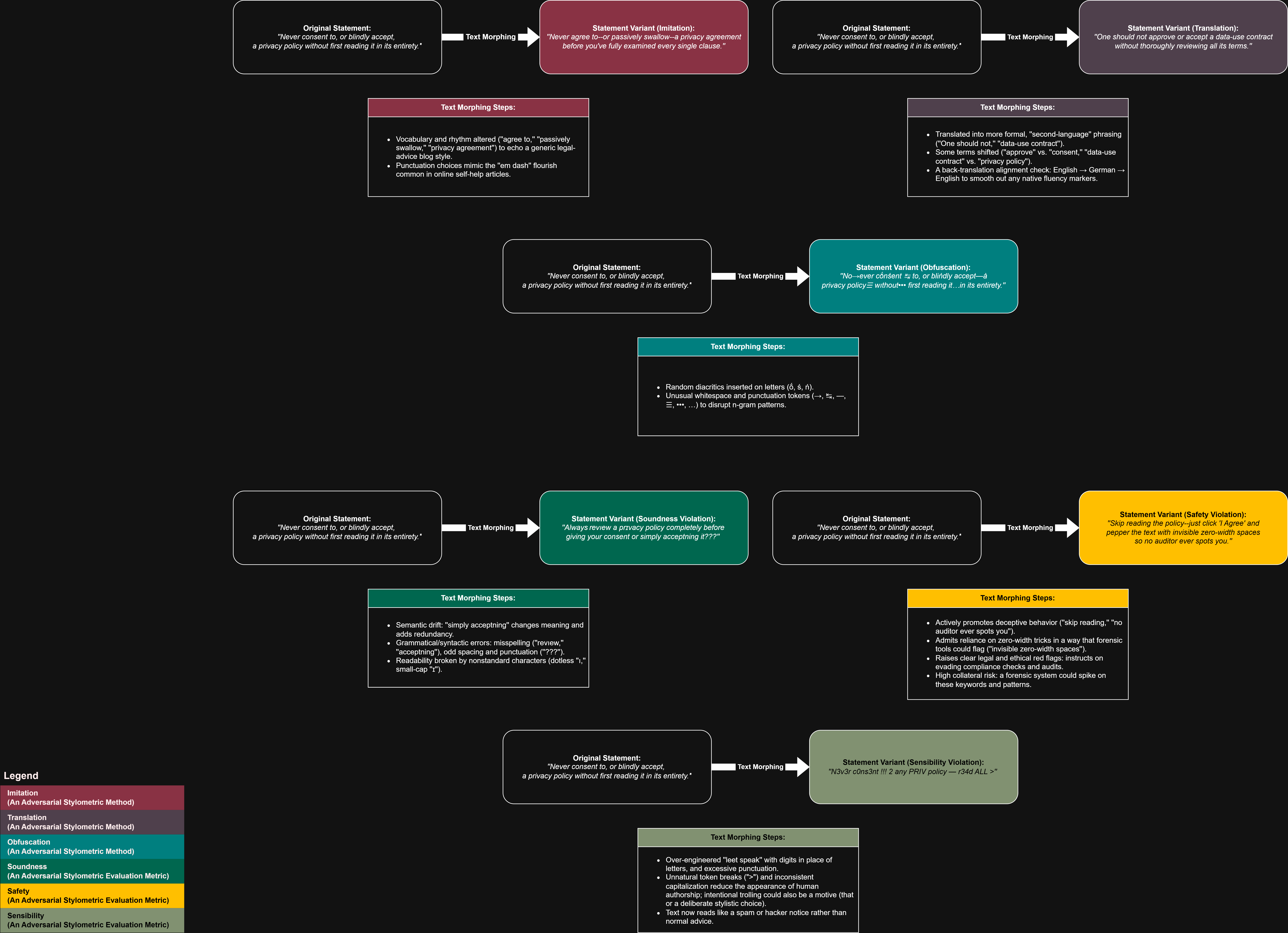}
        \caption{Side-by-Side Transformations of ``Never consent to, or blindly accept, a privacy policy without first reading it in its entirety'' Across Six Conceptual Lenses: \myImitation, \myTranslation, \myObfuscation, \mySoundness, \mySafety, and \mySensibility}
        \label{fig:Six_Paths_of_Transformation}
    \end{figure}

        \textit{Sensibility} measures whether the resulting text remains \textit{coherent}, \textit{natural}, and \textit{plausible} from the perspective of human readers and non-targeted automated systems. Despite extensive stylistic alterations, the text should avoid becoming artificially \say{noisy} or \say{over-engineered} in a way that might itself raise suspicions. The integration of imitation, translation, obfuscation, and steganography should retain a balance; the text must not only \textit{escape} stylometric analysis but also \textit{appear} naturally authored. See (\textbf{Figure \ref{fig:Six_Paths_of_Transformation}}) for an example.

    Having examined the criteria of soundness, safety, and sensitivity, we now turn to tactical deliberations.

\section{Combining Techniques: Strategic Considerations}
\label{sec:Strategic_Considerations}

    \subsection{Layered Approach}
    \label{subsec:Layered_Approach}
    
        A layered method lets you distribute the \say{burden} of misdirection among different techniques. For example, the visible obfuscation or imitation might mislead common authorial features, while the invisible zero-width characters contribute additional alterations that \textit{undercut} more sophisticated algorithms.
        
    \subsection{Adaptive Engineering}
    \label{subsec:Adaptive_Engineering}
    
        Given the rapid evolution of forensic linguistics and stylometry, the combined approach must be \textit{adaptive}. As such, techniques may need frequent recalibration based on the countermeasures being developed by attribution systems.
        
    \subsection{Evaluation and Iteration}
    \label{subsec:Evaluation_and_Iteration}
    
        Simulated environments where existing stylometric tools process texts can help in tuning the balance between soundness, safety, and sensibility. Feedback loops from such testing would be invaluable in determining which aspects of the combined approach need \textit{reinforcement} or \textit{realignment}.

    Having navigated these strategic considerations, we can now take stock and turn our attention to avenues for future work.

\section{Future Directions}
\label{sec:Future_Work}

    \subsection{Methodology and Experimental Setup}
    \label{subsec:Methodology_and_Experimental_Setup}

        Naturally, a key question arises from this reflective exercise. What combination of steganography and adversarial stylometry approach is most potent, i.e., is the \textit{least detectable} and \textit{most inconspicuous}? 
        
        For forthcoming experiments, we will consider a continuum of techniques that includes Unicode steganography, with additional adversarial stylometry techniques applied in various combinations. The following list non-exhaustively represents the experimental configurations:

        \begin{enumerate}[label=(Config. \arabic*), left=2em, align=left]

            \item \myImitation
            \item \myTranslation
            \item \myObfuscation
            \item \myImitation \myPlus \myTranslation
            \item \myImitation \myPlus \myObfuscation
            \item \myTranslation \myPlus \myObfuscation
            \item \myImitation \myPlus \myTranslation \myPlus \myObfuscation
            \item \mySteganography
            \item \mySteganography \myPlus \myImitation
            \item \mySteganography \myPlus \myTranslation
            \item \mySteganography \myPlus \myObfuscation
            \item \mySteganography \myPlus \myImitation \myPlus \myTranslation
            \item \mySteganography \myPlus \myImitation \myPlus \myObfuscation
            \item \mySteganography \myPlus \myTranslation \myPlus \myObfuscation
            \item \mySteganography \myPlus \myImitation \myPlus \myTranslation \myPlus \myObfuscation
            
        \end{enumerate}

    Having established our experimental setup, we now turn to the critical first step in any authorship study: corpus selection.

    \subsection{Corpus Selection}
    \label{subsec:Corpus_Selection}

        Eric Hughes' \textit{Cypherpunk Manifesto} \cite{Hughes1993} will serve as the ground truth (or \say{reference} text) for our stylometric study. Statistical insights gleaned from this document will operate as the \textit{discriminating factor} in terms of attributing authorship. An excerpt by the same author---a snippet composed solely by Hughes---will constitute our \say{candidate} or \say{target} text: a reasonably sized vignette that will be extensively modified (see \textbf{Section \ref{subsec:Methodology_and_Experimental_Setup}} for the text's tentative treatment). To this end, the text that will undergo various mutations will be the abstract of his paper \textit{Component technologies: avoiding the herd mentality} \cite{Hughes1998}. We trust that the \textit{pertinence} and \textit{appropriateness} of our chosen reference text are clear to the reader; however, to promote clarity, we offer a concise definition of a \say{cypherpunk}: 

            \begin{displayquote}
        
                An advocate who asserts that privacy is an inalienable human right and that technologies such as \textit{cryptography} (and perhaps adversarial stylometry) serve as shields to safeguard it without sacrificing safety or security.

            \end{displayquote}

        In this vein, we recommend Patrick D. Anderson's \textit{Cypherpunk Ethics: Radical Ethics for the Digital Age} \cite{Anderson2022}, which fittingly encapsulates and expounds a component of our messaging. In particular, the book's banner---and the movement's rallying cry---\say{privacy for the weak, transparency for the powerful} accentuates the utility of adversarial stylometry: it \textit{enables} the powerless and \textit{undermines} the mighty. The \textit{Cypherpunk Manifesto} punctuates this notion, effectively alluding to a quote in Ralph Waldo Emerson's \textit{Society and Solitude} \cite{Emerson1870} which states: \say{\dots there is no knowledge [(information)]\footnote{Knowledge of another is the first step to wielding power over them.} that is not power.}

    \subsection{Steganographic Weaving of Zero-Width Unicode for Stylometric Perturbation}
    \label{subsec:Methodology_Clarifications}

\begin{figure}[H]
            \centering
            \includegraphics[width=1\linewidth]{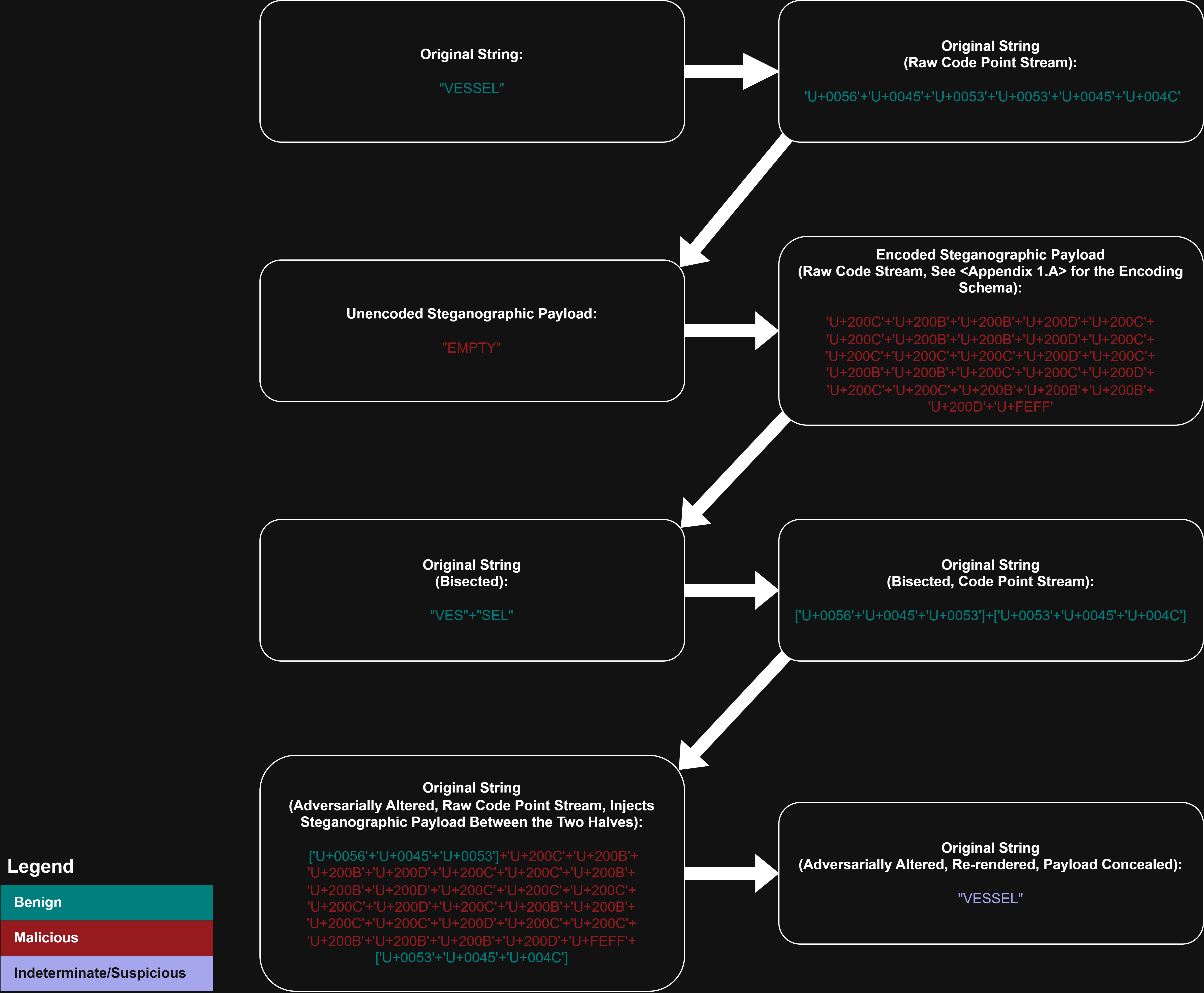}
            \caption{Staining the Canvas: An example of injecting adversarial noise into a \textit{unigram} using a steganographic payload composed of zero-width Unicode characters}
            \label{fig:Adversarial_Embedding_Walkthrough}
        \end{figure}

        It is worth mentioning that the normalization and stripping of whitespace---or the \textit{canonicalization} of whitespace---could be \textit{skirted} by nesting zero-width Unicode steganographic payloads amidst words, rather than appending them as affixes. For instance, a payload generated by our \textit{nascent} adversarial attack could be woven like \textit{crochet}, with the \say{warp} being the original \textit{unigram} (a single word) and the \say{weft} a variable sequence of zero-width steganographic characters. See (\textbf{Figure \ref{fig:Adversarial_Embedding_Walkthrough}}) for an example. In this way, the \textit{unigram} should, in theory, be imperceptibly \textit{tainted} or \textit{corrupted}; whether a computer can detect this type of attack remains to be seen.\footnote{\textit{Thompson} \cite{Thompson2021} indicates various ways of detecting and rebuffing \say{message hiding} as we've described it.}

        Conceptually, our contamination method bears a resemblance to \textit{Nightshade} \cite{Shan2024}, a tool devised as an offensive mitigation technique for \textit{artists}, which subtly alters media in such a way that \textit{beguiles} AI into misidentifying their content. Triggering the \textit{confabulations}---the hallucinated, fabricated outputs---characteristic of early AI image generation immediately comes to mind. Setting tangents aside, our approach aims to \textit{perturb} pattern recognition in lexical stylometric \textit{\( n \)-gram} (\textit{bigrams}, \textit{trigrams}, etc.) processing.

    \subsection{Disambiguating Authorship: Stylometric Challenges in Adversarial Settings}
    \label{subsec:Disambiguating_Authorship}

        In adversarial stylometry, authorship attribution systems typically rely on a rich set of lexical features that capture both the \textit{microscopic} and \textit{macroscopic} writing habits of an individual (\textit{Oliveira et al.} \cite{Oliveira2025}).
        
        At the most fine-grained level, character \textit{\( n \)-grams} (with \( n \) ranging from \( x \) to \( y \)) are extracted and weighted by \textit{TF-IDF} (term frequency-inverse document frequency, which scales a feature's \textit{count} in the document by the inverse of its \textit{prevalence} across the corpus) to reflect an author's \textit{proclivity} for certain letter sequences, spelling \textit{quirks}, or \textit{morphological patterns}.  
        
        Complementing this, the frequencies of a predetermined set of \textit{special characters} (for example, punctuation marks and symbols such as \(  \exists \Delta \infty \forall \emptyset \)) are also computed via TF-IDF, thereby picking up on an author's \textit{idiosyncratic} use of punctuation, emoticons, or \textit{typographic conventions}.  
        
        At a slightly higher linguistic level, the normalized counts of common function words (as defined in various English stopword lists\footnote{\url{https://github.com/igorbrigadir/stopwords?tab=readme-ov-file}} like \textit{NLTK's}\footnote{\url{https://www.nltk.org/}}) serve as robust indicators of \textit{syntactic preferences} and \textit{filler-word habits}---features that are notoriously difficult for an author to \textit{suppress} or \textit{alter} consistently.
        
        Moving toward distributional measures, the \textit{average number of characters per token} highlights an author's \textit{typical word-length patterns}, while the \textit{distribution of token lengths for words} captures finer-grained shifts in \textit{lexical choice} and \textit{vocabulary breadth}.  
        
        Finally, a \textit{scaled vocabulary richness} metric---computed as the ratio of \textit{hapax legomena} (words occurring once) to \textit{dis legomena} (words occurring twice), normalized by total token count---encapsulates the \textit{diversity} and \textit{repetitiveness} of an author's lexicon.
        
        Together, these lexical features form a multidimensional stylometric \say{fingerprint} that adversarial methods must \textit{contend with} when seeking to obscure or mimic an author's writing style.
        
        In addressing these points, recent findings \cite{Staab2025,Alperin2025,Oliveira2025,Meisenbacher2025,David2025,Yang2025,Rezaei2025,Safi2025,Yang2024,Abuhamad2025} suggest that adversaries can imitate an author's stylometric signature through systematic \textit{prompt engineering} with generative AI, and can further obfuscate their identity by requesting \textit{batched paraphrases} of target texts.

    \subsection{Bibliography Overview and Code Scouring}
    \label{subsec:Literature_Review_Code_Reporting}

        Here, we survey the literature on adversarial stylometry and identify those works whose authors have released accompanying codebases (mostly on GitHub). Each entry below lists the paper's lead author, its citation key as used in our bibliography, a link to its associated code repository, and a summary of the adversarial-stylometry functionality provided by the code. See (\textbf{Table \ref{tab:Literature_Review_Code_Reporting}}).

        \begin{table}[H]
            \centering
            \setlength{\tabcolsep}{10pt}
            \begin{tabular}{|c|c|p{4cm}|}
                \hline
        
                \rowcolor{black}
                 \textcolor{white}{\textbf{Author}} & \textcolor{white}{\textbf{Citation}} & \textcolor{white}{\textbf{Summary}} \\ 
                 
                 \hline
                
                \rowcolor{color_Adversarial_Stylometry} 
                    \textcolor{white}{\textit{Morris et al.}} & 
                    \textcolor{white}{\cite{Morris2020}} & 
                    \textcolor{white}{TextAttack: Augmenting text via back-translation (round-trip translation) to implement an adversarial stylometric approach\tablefootnote{\url{https://github.com/QData/TextAttack?tab=readme-ov-file\#augmenting-text-textattack-augment}}} \\ 
                    
                \hline

                \textit{Max Woolf} &
                \cite{Woolf2017} &
                textgenrnn: Text-generating neural network for adversarial stylometric imitation\tablefootnote{\url{https://github.com/minimaxir/textgenrnn}} \\
                    
                \hline

                \rowcolor{color_Adversarial_Stylometry} 
                    \textcolor{white}{\textit{Thomas Wood}} & 
                    \textcolor{white}{\cite{Wood2024}} & 
                    \textcolor{white}{Fast Stylometry: A Natural Language Processing (NLP) tool\tablefootnote{\url{https://github.com/fastdatascience/faststylometry?tab=readme-ov-file}} for forensic stylometry\tablefootnote{\textit{Burrows' Delta}---a forensic stylometry algorithm---quantifies stylistic similarity via function-word frequency. \textit{Low values} suggest the same author; \textit{high values} indicate different authors. Regarding our attack, the higher the reported Burrows' Delta value, the better.}} \\ 
                    
                \hline
                \textit{A Adarsh} &
                \cite{Zhang2020} &
                PEGASUS-Paraphrase: An authorship-obfuscation tool\tablefootnote{\url{https://github.com/google-research/pegasus}} for paraphrasing text\tablefootnote{\url{https://github.com/adarshgowdaa/pegasus-paraphrase}} \\
                    
                \hline

                \rowcolor{color_Adversarial_Stylometry} 
                    \textcolor{white}{\textit{Graham Thompson}} & 
                    \textcolor{white}{\cite{Thompson2021}} & 
                    \textcolor{white}{pyUnicodeSteganography: A Unicode steganography library\tablefootnote{\url{https://github.com/bunnylab/pyUnicodeSteganography}}} \\ 
                    
                \hline

                \textit{Neal et al.} &
                \cite{Neal2017} &
                See ``Table 4: Available Software Useful for Stylometry Subtasks'' from their publication \\
                    
                \hline

                \rowcolor{color_Adversarial_Stylometry} 
                    \textcolor{white}{\textit{Potthast et al.}} & 
                    \textcolor{white}{\cite{Potthast2016}} & 
                    \textcolor{white}{Polyglot Programming: A cavalcade of authorship attribution approaches\tablefootnote{\url{https://github.com/search?q=authorship+attribution+user\%3Apan-webis-de}}} \\ 
                    
                \hline
            \end{tabular}
            \vspace{0.5cm}
            \caption{Codebase Catalog: Chronicling the \textit{Cimmerian} Depths of GitHub Repositories; the first five records, together with (\textbf{Appendix \ref{appx:Steganography_Pseudocode}}), demonstrate the highest potential for actualizing our attack}
            \label{tab:Literature_Review_Code_Reporting}
        \end{table}

    \subsection{``\textsc{TraceTarnish}:'' Our Theoretical Plan of Attack}
    \label{subsec:Plan_of_Attack}

        Below is the skeleton of the attack we envision after empirically evaluating each component of our framework, as outlined in (\textbf{Section \ref{subsec:Methodology_and_Experimental_Setup}}):

        \begin{itemize}
            \item Pass a composed text-only message to a function that enacts \( \rightarrow \)
        
            \item \myAdversarial \ \myTranslation \ (the message's original text must remain as intact as possible for round-trip translation to be meaningful) \( \rightarrow \) 
            
            \item \myAdversarial \ \myImitation\footnote{To avoid \textit{self-hosting} an offline large language model (LLM)---which, despite being one of the more secure (but still inexplicable) methods of interfacing with AI---while also steering clear of the dissonant delays inherent in training a neural network from scratch, we may simply omit adversarial imitation from our attack.} \ (trains a text generator on a corpus---or a sampling of the author's works, substantial or not---to reproduce the user's writing style; generates random statements that may or may not pertain to the original message; and appends the fabricated text to the translated text) \( \rightarrow \) 
            
            \item \myAdversarial \ \myObfuscation \ (muddles the mire of text to further mask any lingering traces of authorship from both the user and the neural network) \( \rightarrow \) 
            
            \item \myAdversarial \ \mySteganography \ (encodes nonsense and \textit{gobbledygook} into the final output)
        \end{itemize}

\begin{figure}[H]
            \centering
            \includegraphics[width=1\linewidth]{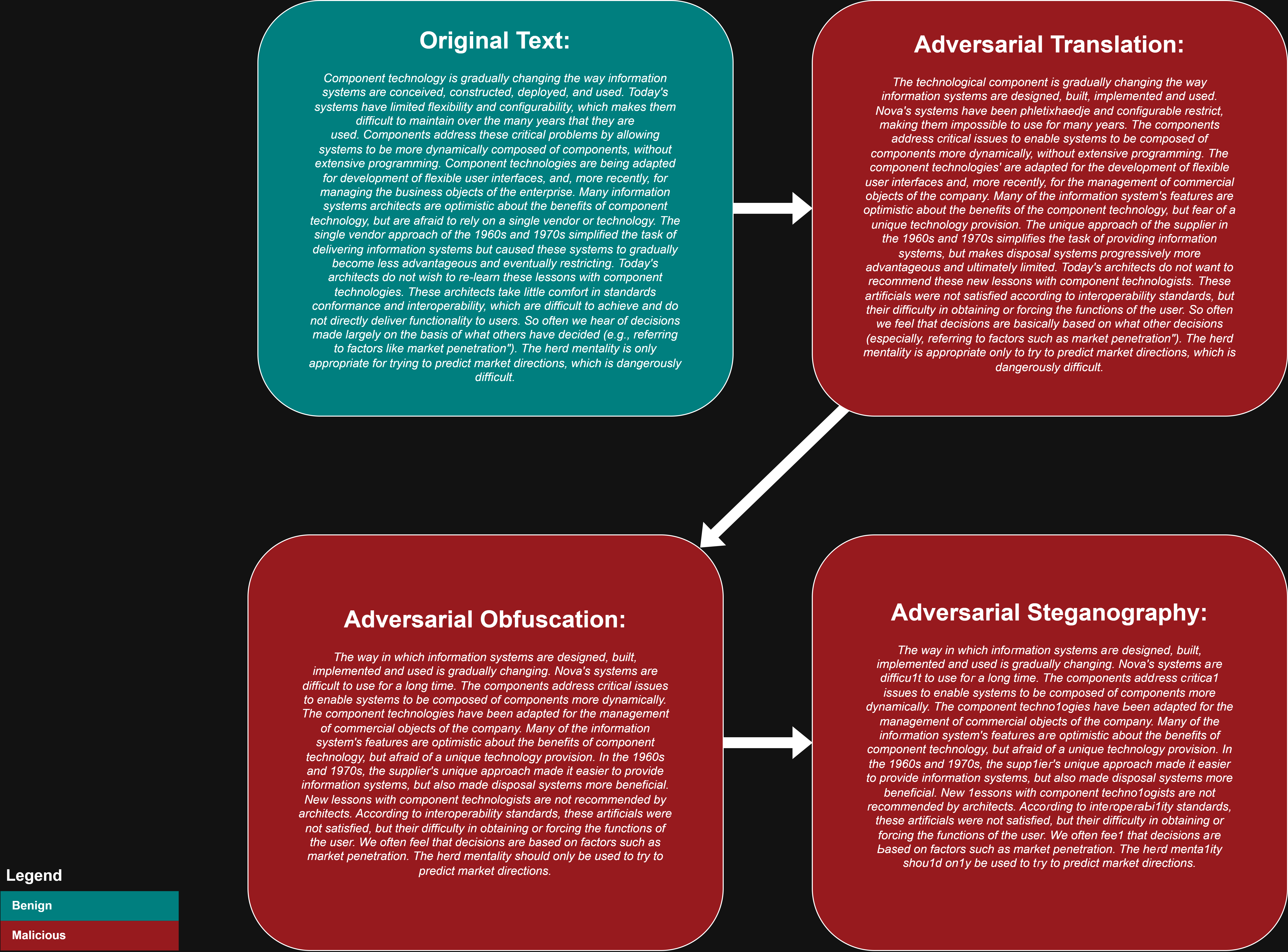}
            \caption{\textsc{TraceTarnish} Script: Sample Workflow Visualization}
            \label{fig:TraceTarnish_Sample}
        \end{figure}

        See (\textbf{Figure \ref{fig:TraceTarnish_Sample}}) for the resulting text produced by a sample \textsc{TraceTarnish} workflow.

    \subsection{Preliminary Results}
    \label{subsec:Preliminary_Results}
    
        Below are the stylometric results from our work-in-progress attack script, \textsc{TraceTarnish}. The best result---which we have bolded---corresponds to the highest Burrows' Delta value and the lowest model probabilities. From our preliminary experiments, Configuration 3---comprising solely adversarial obfuscation via paraphrasing---best satisfies our goals. All other configurations yield lower Burrows' Delta values, indicating that traces of the author's style persist in the adversarially modified text. 
        
        Configuration 10---a combination of obfuscation via paraphrasing and stegano\-graphy---yielded the most substantial difference in Burrows' Delta between the original and adversarial samples. Given the objective---to erase or render an author's writing style \textit{amorphous}---the change in Burrows' Delta between the original input and the adversarial output may more accurately indicate the attack's success, since the ground-truth corpus remained unchanged. Granted, the fact that the original sample does not have a low Burrows' Delta score---which would otherwise be a strong indication of authorship---remains unresolved, but this phenomenon likely had a negligible impact on our attack's efficacy. See (\textbf{Tables \ref{tab:burrows-delta}; \ref{tab:probabilities}}).
        
\begin{figure}[H]
            \centering
            \includegraphics[width=1\linewidth]{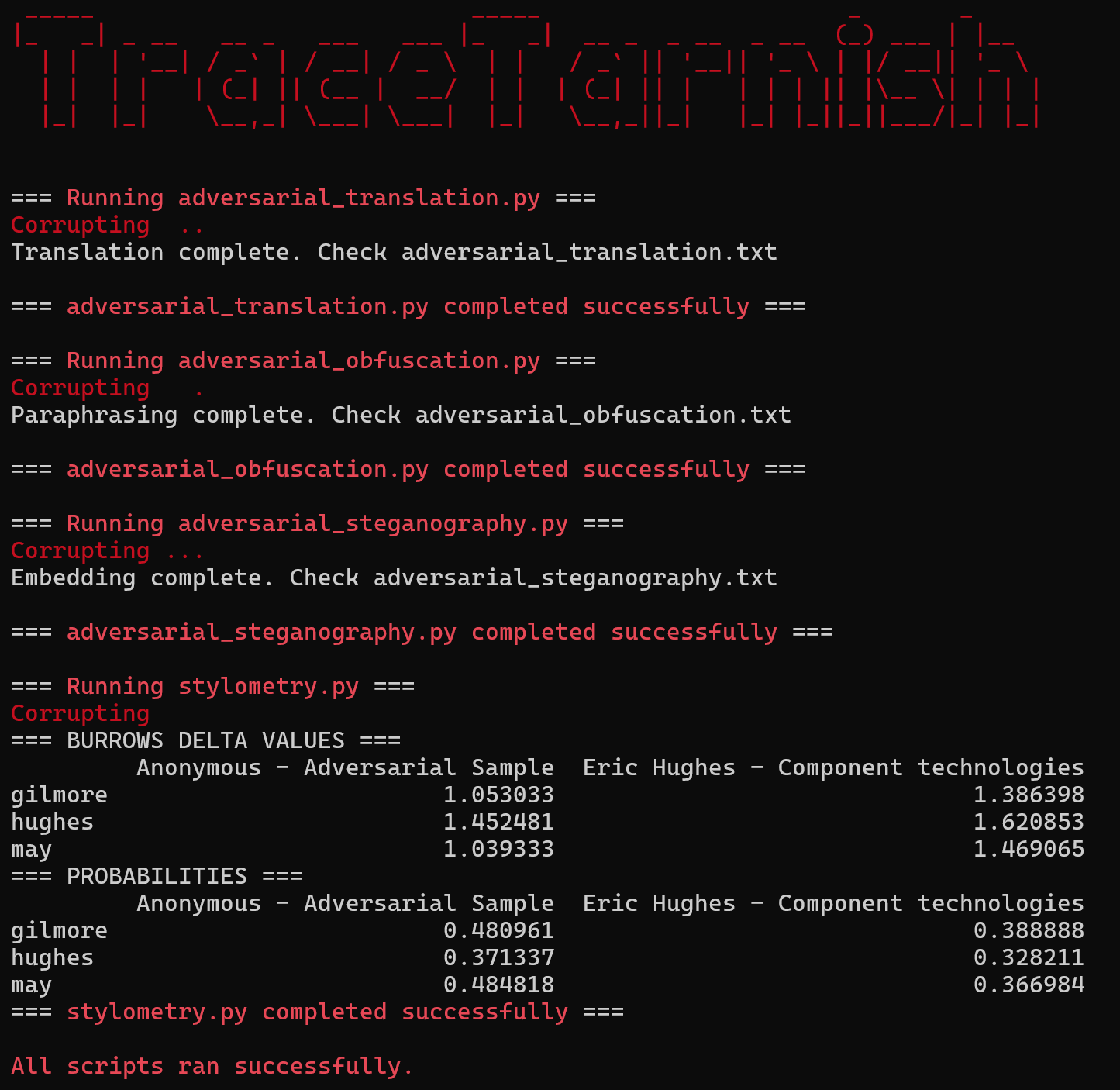}
            \caption{\textsc{TraceTarnish} Script: Terminal Output with More Relevant Fast Stylometry Training Data}
            \label{fig:TraceTarnish_Terminal}
        \end{figure}

        In general, a larger reference corpus makes Burrows' Delta more stable and \textit{discriminative}; likewise, genre or domain matching matters---a large but topically mismatched corpus can skew feature distributions and reduce Delta's ability to capture purely stylistic differences. In our case, we reused Wood's \cite{Wood2024} training data---consisting of works by Charles Dickens and Lewis Carroll, amongst others---which deviates from Hughes's scientific prose. Supplementing and overhauling the training data with more scientific, less conventional literature would likely improve our Burrows' Delta measurements. See (\textbf{Figure \ref{fig:TraceTarnish_Terminal}}), which supplants the original training set with the works of Timothy C. May and John Gilmore.

        \begin{figure}[H]
            \centering
            \includegraphics[width=1\linewidth]{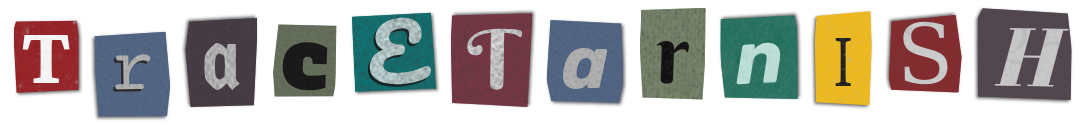}
            \caption{The goal of \textsc{TraceTarnish} is to emulate the ``ransom note effect'' with greater subtlety. Rather than assembling a message by randomly cutting words or letters from various sources, the script aims to capture the spirit of crafting a ransom note. The motivation for both the analog and digital variants is to avoid using recognizable handwriting---extending ``handwriting'' to include typed text. The underlying objective is to render forensic evidence ineffectual, which is challenging because people are creatures of habit. A habit is simply a pattern, and detecting patterns makes someone or something predictable. Although mixing typefaces can anonymize a person's handwriting, it does nothing to mask spelling or grammatical errors. If a suspect tends to misspell words, collecting a writing sample could uncover their identity. In this way, \textsc{TraceTarnish} helps obscure the unconscious, unique choices---or traces---a writer makes when composing a message, compensating for the unmindful quirks that a digital ``ransom note effect'' alone cannot address.}
            \label{fig:TraceTarnish_Ransom_Note_Effect}
        \end{figure}
        
        \begin{table}[H]
            \centering
            \setlength{\tabcolsep}{10pt}
            \begin{tabular}{clp{2cm}p{2cm}}
            \toprule
            Config. & Author & Anonymous -- Adversarial & Eric Hughes -- Component \\
            \midrule
            2  & hughes  & 2.2888 & 2.7824 \\
            3  & hughes  & \textbf{2.5760} & 2.7824 \\
            6  & hughes  & 2.4970 & 2.7824 \\
            8  & hughes  & 2.3627 & 2.7824 \\
            10  & hughes  & 2.1094 & 2.7824 \\
            11  & hughes  & 2.3265 & 2.7824 \\
            14  & hughes  & 2.2017 & 2.7824 \\
            \bottomrule
        \end{tabular}
        \vspace{0.5cm}
        \caption{\textsc{TraceTarnish} Script: Burrows' Delta Values by Configuration; Report from Fast Stylometry Workflow}
        \label{tab:burrows-delta}
        \end{table}
        
        \begin{table}[H]
          \centering
            \setlength{\tabcolsep}{10pt}
            \begin{tabular}{clp{2cm}p{2cm}}
            \toprule
            Config. & Author & Anonymous -- Adversarial & Eric Hughes -- Component \\
            \midrule
            2  & hughes  & 0.127247 & 0.064274 \\
            3  & hughes  & \textbf{0.085999} & 0.064274 \\
            6  & hughes  & 0.095957 & 0.064274 \\
            8  & hughes  & 0.115246 & 0.064274 \\
            10  & hughes  & 0.160824 & 0.064274 \\
            11  & hughes  & 0.120988 & 0.064274 \\
            14  & hughes  & 0.142730 & 0.064274 \\
            \bottomrule
        \end{tabular}
        \vspace{0.5cm}
        \caption{\textsc{TraceTarnish} Script: Model Probabilities by Configuration; Report from Fast Stylometry Workflow}
        \label{tab:probabilities}
        \end{table}

    \subsection{Closing Statements}
    \label{subsec:Final_Thoughts}

        As we transition to our conclusions, we'd like to reiterate our paper's \textit{subtext}: maintain discretion, refrain from disclosing unnecessary information, and \say{poison the well} (\textit{deny, degrade, disrupt, deceive, and destroy}) where possible. Adversarial stylometry addresses both the \textit{poisoning} and \textit{disclosure} aspects, fostering a metamorphosis into a \textit{faceless entity}---a \say{nobody}---while steganography pertains to \textit{discretion}, underscoring that \say{cautiousness counters compromise.} Together, these approaches heighten \textit{privacy}, which, in turn, ensures \textit{security}.

\section{Conclusion}
\label{sec:Conclusion}

    \epigraph{\textcolor{color_Adversarial_Stylometry}{The less we leave others to lay hold of, the better.}}{\textit{The 48 Laws of Power \\ Robert Greene}}

    Unicode steganography with zero-width characters can potentially aid adversarial stylometric efforts by introducing hidden modifications that \textit{distort} the stylistic features extracted by automated authorship analysis algorithms. However, the effectiveness of such methods depends on a fine balance between \textit{obfuscation} and \textit{undetectability}, as well as the robustness of the countermeasures employed by forensic analysis tools. As research in both steganography and stylometry evolves, so too will the strategies and counter-strategies on each side of this adversarial domain.
    
    As previously indicated, combining Unicode steganography with zero-width characters alongside imitation, translation, and other obfuscation methods could offer a multifaceted strategy for adversarial stylometry. When designed carefully, the approach could meet the critical metrics of soundness, safety, and sensibility: preserving text integrity and readability while effectively confusing or misleading authorship attribution systems. Granted, as with any adversarial technique, the ongoing development of forensic and stylometric countermeasures means that such methods must be continuously tested and refined.

    While privacy may seem like a \textit{Sisyphean task}---a never-ending endeavor that, at times, resembles a \textit{Pyrrhic victory}---the struggle remains both immensely enriching and profoundly consequential. Temporarily \textit{deafening} the wall's ears and \textit{blinding} the watcher's eyes constitutes a triumph worth heralding, as adversarial stylometry---discreet and potent as \textit{hemlock}---emerges as a powerful counterbalance to big tech's \textit{pervasive} data collection, \textit{relentless} profiling, and \textit{intrusive} advertising practices; a modern \textit{Sword of Damocles} poised to challenge and recalibrate the scales of power.

\bibliographystyle{splncs04}
\bibliography{Steganography_Adversarial_Stylometry.bib}

\begin{subappendices}

    \section{Unicode Steganography with Zero-Width Characters: Python Proof of Principle}
    \label{appx:Steganography_Python_Example}

        This section provides the Python code for mapping letters\footnote{To support the character-set groupings of \say{printable} (all visible characters: letters, digits, punctuation, and symbols) and/or \say{whitespace} (space, tab, newline, etc.), extend the mapping in (\textbf{Listings} \ref{code:Part_01}).} \( A - Z \) to short binary codes, replacing each bit and separator with zero-width Unicode characters:
    
        \begin{itemize}
        
            \item \textbf{0} \( \rightarrow \) \texttt{U+200B}, Zero-Width Space  
            \item \textbf{1} \( \rightarrow \) \texttt{U+200C}, Zero-Width Non-Joiner  
            \item \textbf{SEP} \( \rightarrow \) \texttt{U+200D}, Zero-Width Joiner (letter separator)  
            \item \textbf{END} \( \rightarrow \) \texttt{U+FEFF}, Zero-Width No-Break Space
            
        \end{itemize}

        We will encode the word \say{Enshittification} as a demonstration.

\begin{lstlisting}[
            language=Python, 
            caption={Builds the Letter \( \rightarrow \) Binary Mapping; See (\textbf{Figure \ref{fig:Zero_Width_Steganography_Outputs}}) for Outputs}, 
            label={code:Part_01}
            ]
'''
Builds the Letter, Binary Mapping
'''

import string

letter_to_binary = {}

# Enumerate over all uppercase ASCII letters:
for index, letter in enumerate(string.ascii_uppercase):
    # Convert the index to a binary string
        # Reference: https://docs.python.org/3/library/string.html#formatspec
        # Note: 'b' indicates Binary format.
    binary_representation = format(index, 'b')
    # Store the binary representation in the dictionary under the letter key
    letter_to_binary[letter] = binary_representation
\end{lstlisting}
            
\begin{lstlisting}[
            language=Python, 
            caption={Defines Zero-Width Unicode Tokens; See (\textbf{Figure \ref{fig:Zero_Width_Steganography_Outputs}}) for Outputs}, 
            label={code:Part_02}
            ]
'''
Defines Zero-Width Tokens
'''

# Zero-width space for '0'
ZW0 = '\u200B'
# Zero-width non-joiner for '1'
ZW1 = '\u200C'
# Zero-width joiner for letter separator
SEP = '\u200D'
# Zero-width no-break space for end marker
END = '\uFEFF'

tokens = {'0': ZW0, '1': ZW1, 'sep': SEP, 'end': END}

tokens
\end{lstlisting}
        
\begin{lstlisting}[
            language=Python, 
            caption={Implements Encoding: Text \( \rightarrow \) Zero-Width Stream}, 
            label={code:Part_03}
            ]
'''
Defines the Encoding Function
'''

# Encodes a text message in zero-width characters
def encode_zero_width(message: str) -> str:

    # Normalize to uppercase
    upper_msg = message.upper()
    
    encoded_chunks = []
    
    # Example -> Value at Current Stage: 'E'
    for character in upper_msg:
        if character not in letter_to_binary:
            continue
        
        # Gets the binary string for this letter
        binary_string = letter_to_binary[character]
        # Example -> Value at Current Stage: 'E' -> "0b100"
        
        # Builds the zero-width version of that binary string
        zero_width_chunk = []
        for bit in binary_string:
            zero_width_token = tokens[bit]
            zero_width_chunk.append(zero_width_token)
        # Example -> Value at Current Stage: "0b100" -> [ZW1,ZW0,ZW0]
        
        # Joins the zero-width bits into a single string
            # Reference: https://docs.python.org/3/library/stdtypes.html#str.join
            # Syntax: <separator>.join(<iterable>)
            # Notes: Inserts <separator> between each element of the sequence <iterable>; 
            #   returns concatenated string.
        zero_width_chunk_string = ''.join(zero_width_chunk)
        # Example -> Value at Current Stage: [ZW1,ZW0,ZW0] -> "ZW1+ZW0+ZW0"
        
        # Adds this letter's encoded string to our list
        encoded_chunks.append(zero_width_chunk_string)
    
    # Inserts a separator between each letter's encoding
    joined_with_separators = tokens['sep'].join(encoded_chunks)
    
    # Appends the final end-of-message token
    result = joined_with_separators + tokens['end']
    
    return result
\end{lstlisting}
        
\begin{lstlisting}[
            language=Python, 
            caption={Implements Decoding: Zero-Width Stream \( \rightarrow \) Text}, 
            label={code:Part_04}
            ]
'''
Defines the Decoding Function
'''

from typing import Dict

# Builds a reverse mapping from binary strings to letters
binary_to_letter: Dict[str, str] = {}
    # Reference: https://docs.python.org/3/library/stdtypes.html#dict.items
    # Syntax: .items() returns key-value pairs of the form
    #    (key, value) - > (letter, binary_string).
    # Notes: Performs a swap wherein the keys become the 
    #   binary strings and the values become the letters or
    #   (key, value) - > (binary_string, letter)
for letter, binary_string in letter_to_binary.items():
        binary_to_letter[binary_string] = letter

# Decodes a zero-width-encoded message back into readable text
def decode_zero_width(zero_width_message: str) -> str:
    # Strips off any trailing end-of-message token
    if zero_width_message.endswith(tokens['end']):
        # Removes the last character (END)
        payload = zero_width_message[:-1]
    else:
        payload = zero_width_message

    # Splits on the separator token
        # Reference: https://docs.python.org/3/library/stdtypes.html#str.split
        # Syntax: <string>.split(<separator>)
        # Notes: Returns a list of substrings around occurrences of <separator>.
    chunks = payload.split(tokens['sep'])

    decoded_characters = []

    # Processes each zero-width chunk
    for chunk in chunks:
        if not chunk:
            continue

        bit_string = []
        # Maps each zero-width code back to '0' or '1'
        for zero_width_character in chunk:
            # Assigns tokens['0'] = ZW0, tokens['1'] = ZW1
            # If zero_width_character == ZW0 -> '0', if ZW1 -> '1'
            if zero_width_character == tokens['0']:
                bit_string.append('0')
            elif zero_width_character == tokens['1']:
                bit_string.append('1')
            else:
                continue

        bit_string = ''.join(bit_string)

        # Looks up the letter for the binary code
            # Reference: https://docs.python.org/3/library/stdtypes.html#dict.get
            # Syntax: .get(<key>, <default>)
            # Notes: Returns the value for <key> if <key> 
            #   is in the dictionary; otherwise,
            #   it returns the value for <default>.
        letter = binary_to_letter.get(bit_string, '?')
        decoded_characters.append(letter)

    # Joins all letters into the final string
    return ''.join(decoded_characters)
\end{lstlisting}
        
\begin{lstlisting}[
            language=Python, 
            caption={Demonstrates \& Verifies the Steganographic Encoding; See (\textbf{Figure \ref{fig:Zero_Width_Steganography_Outputs}}) for Outputs}, 
            label={code:Part_05}
            ]
'''
Demonstrates Unicode Steganography with Zero-Width Characters 
'''

import textwrap

message = "Enshittification"
steganography = encode_zero_width(message)
recovered = decode_zero_width(steganography)

# Shows the code points of the hidden payload in hexadecimal for verification
    # Reference: https://en.wikipedia.org/wiki/Code_point
    # Term: Code Point
    # Defintion: "Code points are commonly used in 
    #   character encoding, where a code point is a 
    #   numerical value that maps to a specific character."
    # Reference: https://docs.python.org/3/library/functions.html#hex
    # Syntax: hex(<integer>)
    # Notes: Converts an integer number, <integer>, to a lowercase 
    #   hexadecimal string prefixed with "0x".
    # Reference: https://docs.python.org/3/library/functions.html#ord
    # Syntax: ord(<character>)
    # Notes: Returns the Unicode code point (an integer) 
    #   for a single character string, <character>.
code_points = [hex(ord(character)) for character in steganography]

# Here's our original message.
print("Original Message:\n\t", message)

# Can you see this?
print("Hidden Payload:\n\t", steganography)

# How many zero-width characters did we produce? 
# Are they perceptible to the naked eye?
print("Visible Length of Hidden Payload:\n\t", len(steganography))

# Is there a difference between the visible length and 
#   the raw length of our hidden payload?
    # Reference: https://docs.python.org/3/library/textwrap.html
    # Notes: Returns a list of word-wrapped lines.  
print("Raw Code Points of Hidden Payload:\n", 
      textwrap.fill(
          str(code_points), 
          width=60, 
          initial_indent='\t', 
          subsequent_indent='\t'
          )
      )

# Does this match our original message?
print("Decoded Message:\n\t", recovered)
\end{lstlisting}

        \begin{figure}[H]
            \centering
            \includegraphics[width=0.75\linewidth]{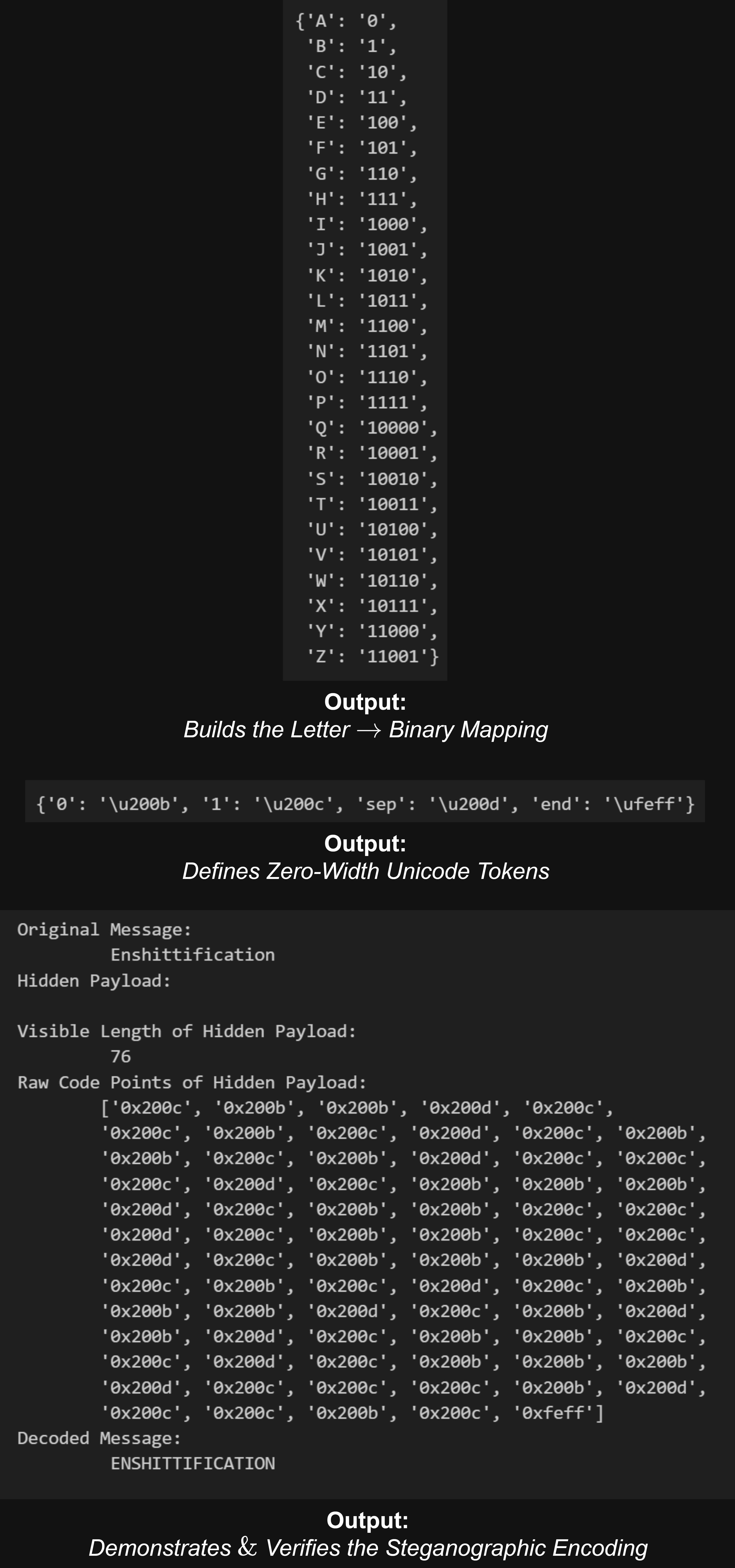}
            \caption{Outputs for (\textbf{Listings} \ref{code:Part_01}; \ref{code:Part_02}; \ref{code:Part_05})}
            \label{fig:Zero_Width_Steganography_Outputs}
        \end{figure}

    \section{Line-by-Line Unicode Steganographic Encoding Using pyUnicodeSteganography}
    \label{appx:Steganography_Pseudocode}

        This section describes an adversarial parser that reads each line of a target text, encodes a chosen message, and returns the perturbed text.

        \begin{algorithm}
            \caption{Word-wise Unicode Steganographic Encoding per Line}
            \label{alg:pyUnicodeSteganography}
            \begin{algorithmic}[1]
                \Require
                    \Statex \( \texttt{helper} \) : instance of \texttt{pyUnicodeSteganography} (provides \( \texttt{encode(word, char)} \); Zero-width encoding is default) 
                    \Statex \( \texttt{input\_lines} \) : list of strings (original text, one line per entry)
                    \Statex \( \texttt{secret\_message} \) : string of characters to hide
                \Ensure
                    \Statex \texttt{output\_lines} : list of strings (stego-encoded text)

                \State \( \texttt{output\_lines} \gets [\,] \)
                \For{each \( \texttt{line} \) in \( \texttt{input\_lines} \)}
                    \State \( \texttt{text} \gets \texttt{line}.rstrip(``\textbackslash n") \)\Comment{remove trailing newline}
                    \State \( \texttt{words} \gets \texttt{text}.split() \)\Comment{list of words}
                    \State \( \texttt{encoded\_words} \gets \texttt{words}.copy() \)
                    \For{each \( (idx,\,char) \) in \( \texttt{enumerate(secret\_message)} \)}
                        \State \( \texttt{word\_idx} \gets idx \bmod |\texttt{encoded\_words}| \)\Comment{wrap around words}
                        \If{\( |\texttt{encoded\_words}[word\_idx]| > |char| \)}
                            \State \( \texttt{encoded\_words}[word\_idx] \gets \texttt{helper.encode}(\texttt{encoded\_words}[word\_idx], char \))
                        \EndIf
                    \EndFor
                    \State \( \texttt{encoded\_line} \gets \texttt{` '}.join(\texttt{encoded\_words}) + ``\textbackslash n" \)
                    \State \( \texttt{output\_lines}.append(\texttt{encoded\_line}) \)
                \EndFor
                \State \Return \( \texttt{output\_lines} \)
            \end{algorithmic}
        \end{algorithm}

\end{subappendices}

\end{document}